\documentclass[proof]{WileyASNA-v1}

\articletype{Article Type}%

\begin{document}

\title{Metal-poor Stars Observed with the Automated Planet Finder Telescope. III. CEMP-no Stars are the Descendant of Population\,III Stars}

\author[1]{Nour Aldein Almusleh}

\author[2]{Ali Taani}

\author[3]{Sergen \"{O}zdemir}

\author[4,5]{Maria Rah}

\author[6,7]{Mashhoor A. Al-Wardat}

\author[8,9]{Gang Zhao}

\author[10]{Mohammad K.\ Mardini*}

\authormark{Nour \textsc{et al}}

\address[1]{Department of Physics, Al al-Bayt University, Al Mafraq, Jordan}

\address[2]{Physics Department, Faculty of Science, Al-Balqa Applied University,Jordan}

\address[3]{Department of Astronomy and Space Sciences, Ege University, 35100 Bornova, \.{I}zmir, Turkey}

\address[4]{National Astronomical Observatories and Key Laboratory of Computational Astrophysics, Chinese Academy of Sciences, 20A Datun Rd., Chaoyang District, Beijing 100012, China}

\address[5]{Research Institute for Astronomy and Astrophysics of Maragha (RIAAM), University of Maragheh, Maragheh, Iran}

\address[6]{Department of Applied Physics and Astronomy, University of Sharjah, Sharjah, United Arab Emirates}

\address[7]{ Sharjah Academy for Astronomy, Space Sciences and Technology, University of Sharjah, Sharjah, United Arab Emirates}

\address[8]{Key Lab of Optical Astronomy, National Astronomical Observatories, Chinese Academy of Sciences, Beijing 100101, China}

\address[9]{Institute of Space Sciences, Shandong University, Weihai 264209, China}
\address[10]{Kavli Institute for the Physics and Mathematics of the Universe (Kavli IPMU, WPI), The University of Tokyo, Chiba 277-8583}

\corres{*Mohammad K.\ Mardini, Kavli Institute for the Physics and Mathematics of the Universe (Kavli IPMU, WPI), The University of Tokyo, Chiba 277-8583. \email{mardini\_mohammad@hotmail.com}}

\abstract{This study reports a probabilistic insight into the stellar-mass and supernovae (SNe) explosion energy of five CEMP-no stars' possible progenitors. As such, a direct comparison between the abundance ratios [X/Fe] of the light-elements and the predicted nucleosynthetic yields of SN of high-mass metal-free stars has been performed. This comparison suggests that the possible progenitors have a stellar-mass range of 11 - 22\,M$_{\odot}$ and explosion energies of $0.3 - 1.8 \times 10^{51}$\,erg. In addition, we investigate the kinematic signatures, derived from $Gaia$ DR2, to conclude that these five CEMP-no stars have halo-like kinematic and do not enter the outer-halo region. In addition, we link the abundance patterns with kinematic signatures to investigate the $Gaia$-Sausage and $Gaia$-Sequoia memberships. This chemo-dynamical analysis, suggests that these CEMP-no stars are not $Gaia$-Sausage nor $Gaia$-Sequoia remnant stars, but another accretion event might be responsible for the contribution of these stars to the Galactic Halo..}


\maketitle

\section{Introduction} \label{sec:intro}

Population\,III (Pop\,III) stars are formed in metal-free gas clouds as the hydrogen molecules dominated the cooling process, shortly after the Big Bang \citep[for a selected list see, e.g.,][]{1969PThPh..42..219M, 1972PASJ...24...87Y, 1976ApJ...205..103H, 1977ApJ...211..638S, 1977ApJ...214..152S, 1979PASJ...31..505Y}. In comparison with nowadays stars, Pop\,III stars are believed to be very massive, very luminous, and short-lived.
This type of stellar objects is believed to be responsible for producing the first metals in the universe through their supernova explosions (SNe). Therefore, investigating the yields of these SNe would allow us to study the chemical nature and physical processes that govern our early Galaxy. \citep{Mardini_2019b, 2020ApJ...897...78P, Mardini_2020}. Unfortunately, these primordial SNe are difficult to observe due to faintness from their large redshifts.

A promising alternative is to connect existing ancient stars Pop\,II (the so-called metal-poor) in the Milky Way to the processes that governed the early universe. This approach is known as \textit{Galactic Archaeology}, in which long-lived metal-poor stars are identified and used as time capsules. Therefore, the science of \textit{Galactic Archaeology} is built on two fundamental points:
i) Pop\,II (the so-called metal-poor) stars preserve in their atmosphere the chemical composition of individual or a few SN yields of previous Pop\,III stars \citep[the so-called Pop III chemical fingerprint,][]{2016ApJ...833...21P, Mardini_2019b, Mardini_2019a,2020ApJ...897...78P} and (ii) the derived chemical abundances from the high-resolution spectroscopic data is reliable to perform accurate comparisons with the SNe yields  \citep[e.g.,][]{Mardini_2020, 2020NatAs...4..564F}.

Pop\,II stars are known to be deficient in iron \citep[e.g.,][]{2011Natur.477...67C,2018MNRAS.481.3838S, 2021arXiv210316660C} and enriched in alpha elements \citep[e.g.,][]{2004A&A...416.1117C,2009A&A...501..519B}. But, some Pop\,II stars show different chemical patterns (e.g., carbon-enhancement, $r$-process enhancement). Metal-poor stars that show a high carbon-to-iron abundance ratio are called carbon-enhanced metal-poor (CEMP; see 
\citealp{2005ARA&A..43..531B, 2007ApJ...655..492A, 2014ApJ...797...21P}; for more details about their classification criteria) stars. Galactic archaeologists are very interested in CEMP stars since they dominate the tail of the metal-poor region (with frequency $\sim$ 80\% at [Fe/H] $< -$4.0; \citealp{2014ApJ...797...21P}) and might hold a diagnostic key to further our understanding of the early universe. The analysis of CEMP stars, in the literature, confirmed the existence of several distinguished heavy-element patterns, namely: i) CEMP-s stars ([Ba/Fe]$>1$, [Ba/Eu]$>0$, and [Eu/Fe]$\leq 1$), which show enrichment in barium (mainly form by \textit{s}-process, see \citealp{2000ApJ...544..302B}), ii) CEMP-r stars ([Eu/Fe]$>1$ and [Ba/Eu]$<0$), which show enrichment in europium (mainly form by \textit{r}-process, see \citealp{2000ApJ...544..302B}), CEMP-s/r ([Ba/Fe]$>1$, [Ba/Eu]$>0$, and [Eu/Fe]$>1$) stars, which show enrichment in both barium and europium, iv) CEMP-no ([Ba/Fe]$<0$) stars, which show no enrichment in heavy elements.

The large fraction of the CEMP-no stars among the most primitive metal-poor stars increase their importance (e.g., \citealp{2014ApJ...797...21P}) and their observed sub-solar heavy elements abundances reflect the composition of their born natal clouds, which got polluted by an individual or a few SNe yields of the previous Pop\,III. Therefore, expanding the chemical inventory and the numbers of well-studied CEMP-no stars will help our understanding of the galactic halo's chemical-evolution.

Spatially, metal-rich stars are mainly found in the Galactic bulge and the Galactic disk. On the other hand, metal-poor stars are found, generally, in the Galactic halo \citep{2021arXiv210316642C}. Although the vast majority of the galactic halo stars might have an extragalactic origin (particularly the outer-halo), it is worth emphasizing that a significant number of in-situ stars reside primarily in the few kpc of the solar neighborhood \citep[e.g.,][]{2009ApJ...702.1058Z,2014MNRAS.439.3128T}. Interestingly, 334 metal-poor stars, with $-3.7 <$ [Fe/H] $<-2.35$, have been assigned to the thick disk (Mardini et al. 2021 in prep). Also, similar results have been reported by other authors \citep[e.g.,][]{2020A&A...636A.115D, 2019MNRAS.484.2166S}. These stars, with disk-like kinematics, suggest that the galactic disk should not be avoided in the process of searching for the most metal-poor stars. In the era of the second data release of the $Gaia$ mission \citep[$Gaia$ DR2;][]{2018A&A...616A...1G}, several studies investigated the kinematics and/or chemical abundance patterns of CEMP stars to gain insight into their putative progenitor (e.g., \citealp{Mardini_2019a}).
In this paper, we perform a chemo-dynamical analysis of five CEMP-no stars based on the $Gaia$ DR2 astrometry, galactic potential, and the state-of-the-arts of the action-based galaxy modeling. We aim to improve our understanding of the CEMP-no stars' origin by linking their chemistry and motions. The results of this work are expected to improve the constraints of the Galactic halo's assembly history.

\section{Observational Data} \label{sec:data}
\subsection{Chemical Abundances}

This is a continuation of the work presented in \citet{Mardini_2019a} and \citet{Mardini_2019b} (hereafter Paper I and II). We refer the reader to Paper I and II for more details about targets selection, chemical abundances' derivations, and the overall scientific goals, which we briefly outline here. In this series of papers, the metal-poor candidates were chosen based on a direct comparison between the observed flux, in the third release of the Large Sky Area Multi-Object Fibre Spectroscopic Telescope (LAMOST; \citealp{Zhao_2006,Zhao_2012,2012RAA....12.1197C}), and the flux of a synthetic grid of spectra. Stars with [Fe/H] $\leq -2.7$ were identified as very metal-poor candidates. For a small subsample of bright stars (13 stars), we obtained high-resolution ($R=110,000$ carried out with a slit width of 0.5\,mm) using the 2.4\,m Automated Planet Finder. Using these high-resolution spectra, in Paper I and Paper II we derived chemical abundances for $\sim$ 25 elements. Five of those 13 stars show relatively high enhancement in carbon abundances, which we use in the following analysis.

In this paper, we adopted the light-elements (Z $\leq 30$) abundances from Paper I and Paper II (see Table~4 in both papers). The adopted elements are sufficient to make a direct comparison with the theoretical yields of SNe. \cite{Hegerwoosley10} computed 16,800 nucleosynthetic gains of SNe of high-mass metal-free stars. The well-known chi-square matching algorithm has been used to determine the best parameter set. The parameter set includes progenitor masses ranging between 10 and 100 solar masses, explosion energies corresponding to those mass intervals, and the mixing factor ($f_{min}$). Further details about the parameter space and the model description can be found in \citet{Hegerwoosley10}.

\subsection{Six-dimensional Phase Space Coordinates}

The positions ($\alpha, \delta$), proper motions ($\mu_{\alpha} cos \delta, \mu_{\delta}$), and parallaxes ($\varpi$) of our targets were taken from $Gaia$ DR2 \citep{2018A&A...616A...1G}. Deriving an accurate stellar distance from an observed parallax is straightforward as $d=1/\varpi$, only when $\varpi > 0$ and $f \equiv \sigma_\varpi/\varpi \lesssim 20 \%$ \footnote{Where $\varpi$ is the observed parallax and $\sigma_\varpi$ is the reported random uncertainty}. However, we adopted the distances derived by the Bayesian approach published in \citep{BJones18}, which agree well with the distances obtained by inverting parallax. Finally, we adopted the measured heliocentric velocities in Paper I and II as the line-of-sight velocities.

By adopting the Galactic potential \texttt{MWPotential2014} \citep{Bovy15} and Action-based galaxy modeling architecture (\texttt{Agama}), we performed a detailed orbital integration, derive total orbital energy ($E=(1/2)v^2 + \Phi(x)$), and calculate the three-dimensional action ($\vec{J}=(J_r, J_\phi, J_z)$) (for more details, see \citealt{2019MNRAS.482.1525V}). In addition, we used the apocentric ($r_{apo}$) and pericentric ($r_{peri}$) radii to define the eccentricity as $e=(r_{apo}-r_{peri})/(r_{apo}+r_{peri})$. We use this information to investigate the orbital evolution of our sample around the Milky-Way.

\section{Discussions and Results} \label{discuss}

The collected data is sufficient to proceed with our investigation. However, employing Monte-Carlo-sampling helps to gain a probabilistic insight into the possible origin of these five CEMP-no stars. Therefore, we used a normal distribution, the observed data (chemical abundances and 6D space coordinates), and associated error measurements to generate 10,000 realizations.

\subsection{Possible Progenitors}

\begin{figure*}[!hptb]
\centerline{\includegraphics[width=600pt,height=50pc]{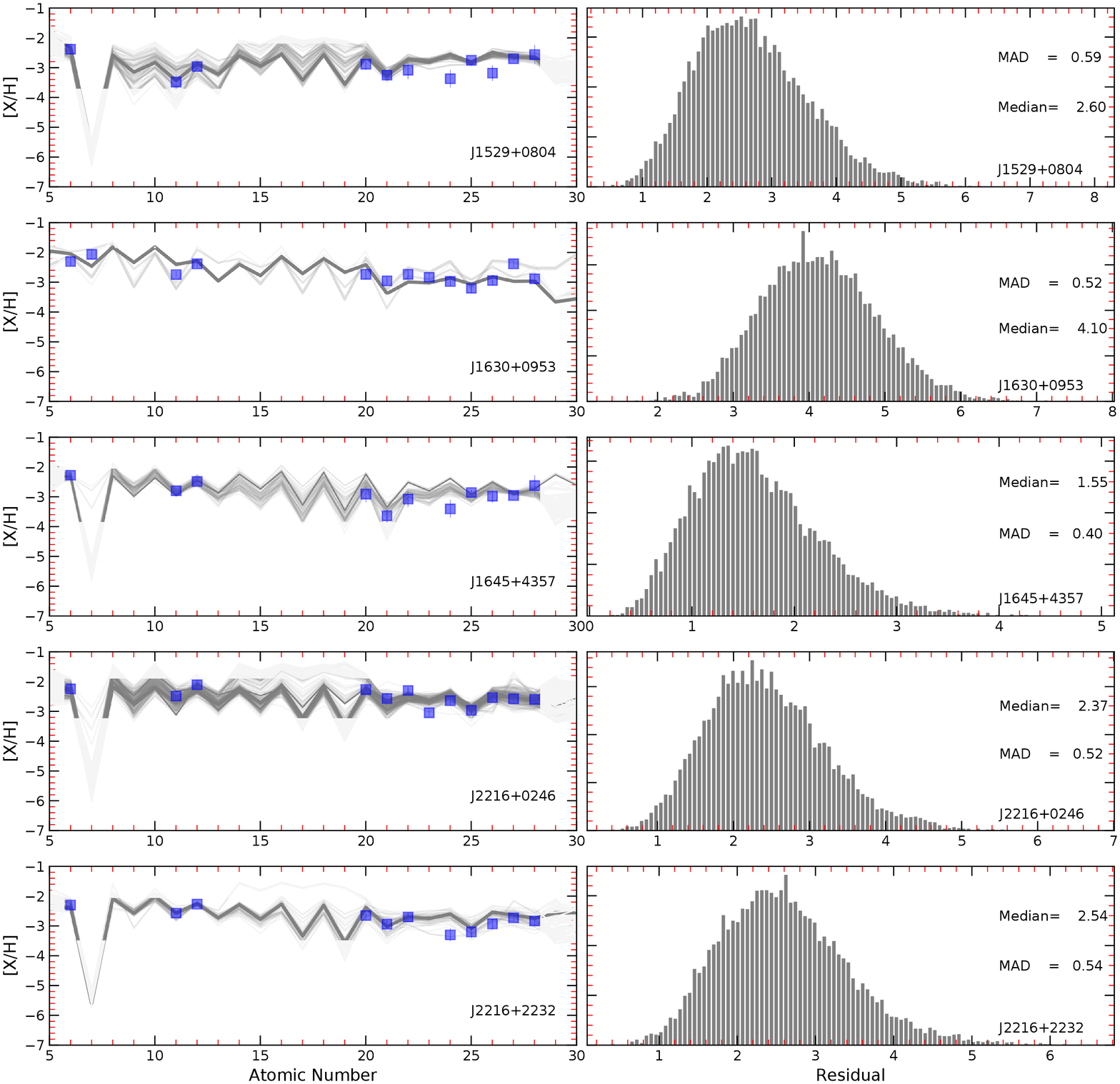}}
\caption{Right panel shows the observed [X/H] abundance ratios of the five CEMP-no stars (filled blue squares) as a function of atomic number, overlaid with the matched SN models. The best fits and their properties are discussed in the text. The left panel shows the posterior distributions for the mean squared residuals $\chi^{2}$, of the 10,000 simulations. The median and median absolute deviation (mad) are shown in legends.\label{fig:fit}}
\end{figure*}

We made a direct comparison between the generated light-element abundances (log $ \epsilon (X)$) and the predicted yields from \citet{Hegerwoosley10}, using the open-source \texttt{STARFIT} code\footnote{http://starfit.org}, to statistically estimate the stellar mass of the possible progenitors of these five CEMP-no stars. It is noteworthy to mention that the likely progenitor of J1630+0953 and J2216+0246 have been discussed in paper II. We illustrate our comparison in Figure~\ref{fig:fit}. On the left panel, we plotted atomic number versus [X/H]\footnote{X refers to different light-elements}. Blue-filled squares represent the adopted chemical abundances, and error bars represent the associated uncertainties. Gray lines show the predicted SN patterns that fit the observed abundance, and their transparency can be used as a means of  pattern repetition. The median value and the median absolute deviation (MAD) are shown in legends. The MAD assesses  how the data spread in the value space.

For the star J1529+0804, we found a model, with a mass of 11.4\,M$_{\odot}$ and a mixing factor of $f_{min}=0.063$, fits 1,561 out of the 10,000 realization. On the other hand, the explosion energy of this possible progenitor is $0.3 \times 10^{51}$\,erg. We estimated more massive progenitors for the rest of our sample 21.5, 20, 13.4, and 15.8\,M$_{\odot}$ for J1630+0953, J1645+4357, J2216+0246, and J2216+2232, respectively. All model details are summarized in Table\ref{tab:models}.

\begin{table*}[!hptb]
\caption{Summary of models for our target stars.}\label{tab:fitting}
\label{tab:models}
\centering
\begin{tabular}{|c|c|c|c|c|c|}
\hline 
\textbf{\rule{0pt}{4ex} Star} & \textbf{Mass} & \textbf{Explosion energy} & \textbf{Mixing Factor} & \textbf{Number Density} & \textbf{Weight} \\ \cline{2-6}
  & M$_{\odot}$ & $\times 10^{51}$ erg & $f_{min}$ & $\bullet$ & $\%$ \\ 
\hline 
J1529+0804 & 11.4 & 0.3 & 0.063 & 1561 & 15.61 \\ \hline 
J1630+0953 & 21.5 & 0.3 & 0.158 & 9140 & 91.40 \\ \hline 
J1645+4357 & 20.0 & 0.3 & 0.158 & 6975 & 69.75 \\ \hline 
J2216+0246 & 13.4 & 1.5 & 0.000 & 2235 & 22.35 \\ \hline 
J2216+2232 & 15.8 & 1.8 & 0.016 & 4362 & 43.62 \\ 
\hline 
\end{tabular} 
\end{table*}

\cite{Ishigaki18} studied about 200 extremely-metal-poor (EMP) stars. The authors showed that the observed abundance patterns in those EMP stars provide some clues about Pop III stars' physical properties and their supernova explosion. They concluded that the vast majority of the progenitors of those EMP stars lie in the stellar-mass range 15-25\,M$_{\odot}$. Our comparison shows that three of our sample stars (J1630+0953, J1645+4357, and J2216+2232) agree well within this stellar mass range. Our results suggest that a single SN ejecta of Pop\,III stars with a stellar-mass range of 11.4-$22.5$\,M$_{\odot}$ was most likely responsible for the observed light elements patterns of these five CEMP-no stars. \texttt{STARFIT} support comparison with multiple SN ejecta, this option did not improve the resulting fitting. Therefore, we argue that no more than one SN (the so-called mono-enrichment) might alter the chemical composition of the birth-cloud of our CEMP-no stars.

\subsection{Orbital Properties}
We considered the known zero-point offset parallax as described in \cite{2018A&A...616A...2L}. We use the sampled astrometric solutions to calculate the Galactocentric Cartesian ($X_{GC}$, $Y_{GC}$, $Z_{GC}$) coordinates as follows:

\begin{center}
$X_{GC} = R_{\odot} - d cos (b) cos(l)$\\
$Y_{GC} = - d cos (b) sin(l)$\\
$Z_{GC} = d sin(b) + z_{\odot}$
\end{center}

The Galactic space-velocity components (\textit{U, V, W}) are calculated and corrected using the Astropy Galactocentric frame package \citep{Astropy13, Astropy18}.  \textit{U} positive toward the Galactic center, \textit{V} positive in the direction of Galactic rotation and \textit{W} positive toward the north Galactic pole. Then we used the angle $\phi = tan^{-1} (Y_{GC}/X_{GC})$ to calculate the cylindrical velocities components for our targets as follows:

\begin{center}
$V_R = U cos(\phi) + V sin(\phi)$\\
$V_{\phi} = U sin (\phi) - V cos(\phi)$\\
$V_Z = W$
\end{center}

Here, we assume the newly determined Solar System of Rest (LSR) and the Sun position and velocities \citep{2010MNRAS.403.1829S,2019A&A...625L..10G, 2019MNRAS.482.1417B,2019AstBu..74..464M, 2019JPhCS1258a2029T,2019AN....340..847T} in the Galactic coordinate frame. The Sun is located at $z_{\odot}$ = 25\,pc, at Galactic distance $R_{\odot}$ = 8.2\,kpc, and with circular velocity $v_{\odot}$ = 232.8\,$\text{kms}^{-1}$. The solar peculiar motion is ($U_{\odot}, V_{\odot}, W_{\odot}$) = (11.1, 12.24, 7.25)\,$\text{km s}^{-1}$.

The orbital properties of J1630+0953 and J2216+0246 have been calculated, in Paper II, using \texttt{MWPotential2014}, while the orbital properties of J1529+0804, J1645+4357, and J2216+2232 were calculated, in Paper I, using St\"{a}ckel potential. It is well known that the \texttt{MWPotential2014} is shallower compared to the St\"{a}ckel one,  resulting in slightly different orbital properties (e.g., the total orbital energy of a star calculated using the St\"{a}ckel will be lower than the orbital energy calculated with the \texttt{MWPotential2014}). Therefore, for the sake of consistency, we re-calculate the orbital properties for the entire sample using the \texttt{MWPotential2014}.

\begin{table*}[!hptb]
\caption{Orbital properties of our sample}
\label{orbital_data}
\centering
\begin{tabular}{|c|c|c|c|c|c|c|c|c|c|c|}
\hline
Star ID& $E\ (\text{km}^{2}/\text{s}^2)$ & $Z_{max}$      (kpc)             & $r_{apo}$ (kpc)       & $r_{peri}$ (kpc)   & $e$   & Distance (kpc) & d1 (kpc)  & Error \\ \hline 
J1529+0804 & $-1.293^{+1.13}_{-0.55}$  & 3.101 & 7.006   & 2.035         & 0.549 & 3.628    & 0.425     & 11.7\% \\ \hline
J1645+4357 & $-1.277^{+1.66}_{-0.87}$  & 3.564 & 8.075   & 0.502         & 0.883 & 8.355    & 0.996     & 11.9\% \\ \hline
J2216+2232 & $-1.109^{+ 0.75}_{-0.69}$ & 8.569 & 9.281   & 4.496         & 0.347 & 2.440    & 0.211     & 8.65\% \\ \hline
J1630+0953 & $-1.188^{+10.33}_{-7.19}$ & 7.149 & 7.165   & 4.201         & 0.304 & 6.183    & 0.83      & 13.4\% \\ \hline
J2216+0246 & $-1.277^{+1.65}_{- 0.89}$ & 3.615 & 8.077   & 0.494         & 0.885 & 2.899    & 0.38      & 13.1\% \\ \hline
\multicolumn{2}{l}{Note: d1  is the 16th percentile}
\end{tabular}
\end{table*}

\begin{figure*}[!hptb]
\centerline{\includegraphics[width=600pt,height=50pc]{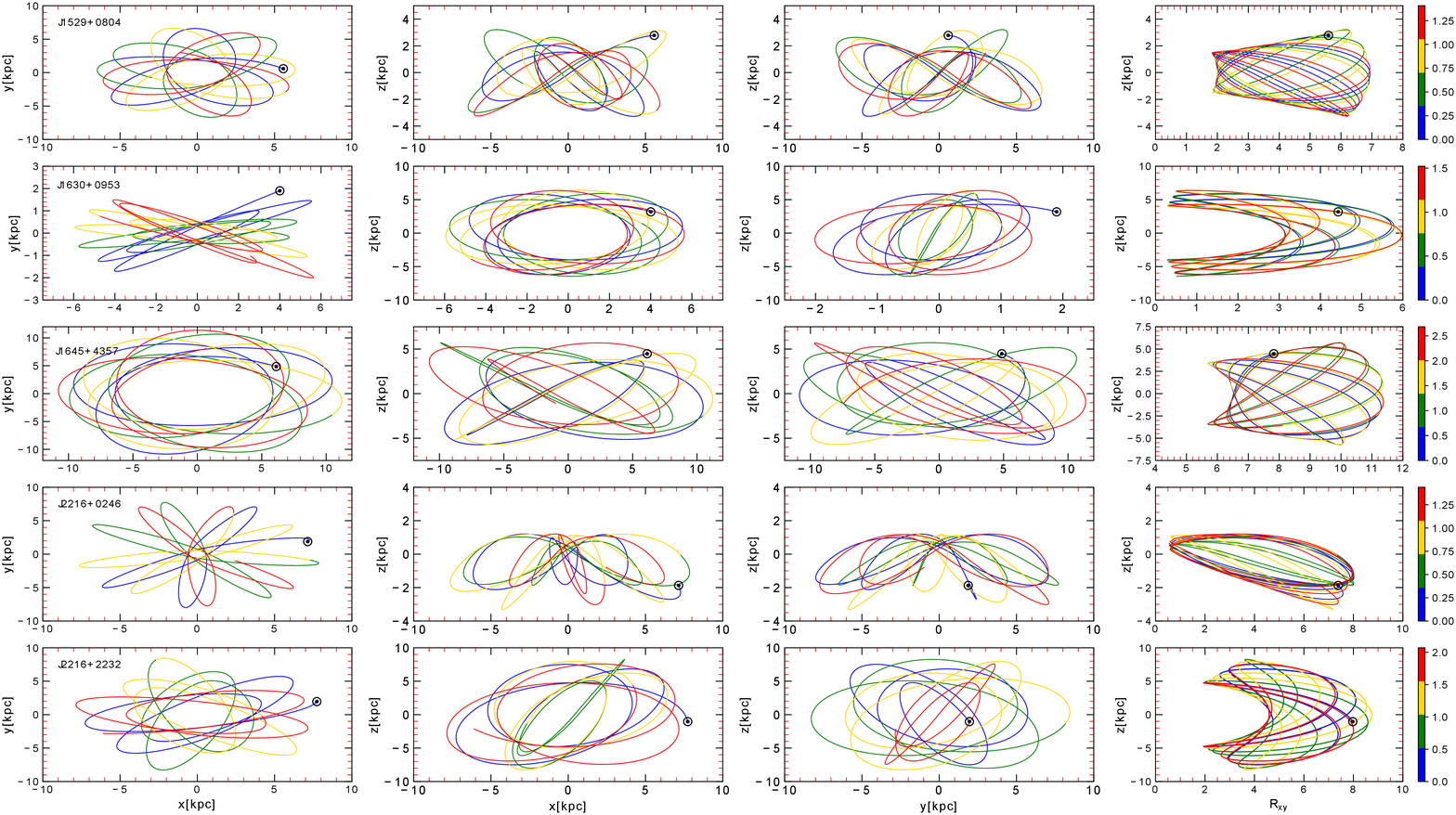}}
\caption{Orbits of our sample stars integrated in time (color-bar) in \texttt{MWPotential2014}. The first and second columns show $Y_{GC}$ and $Z_{GC}$ as a function of $X_{GC}$, respectively. The third and fourth columns show $Z_{GC}$ as a function of $Y_{GC}$ and the galactocentric radius, respectively. The dotted-open circles denotes the current location of our stars in different frames. Each row represent the orbits of one star (see legends).\label{fig:orbits}}
\end{figure*}

Figure~\ref{fig:orbits} shows the last 10 periods of our sample's orbit in different projections integrated in time. Based on the criteria discussed in this series (star with total energy higher than $-0.9\,\text{km}^2 \text{s}^{-2}$ and at distance $r_{apo}$ greater than 15\,kpc) and the orbits of our stars (shown in Figure~\ref{fig:orbits}), we suggest that our stars have halo-like kinematics, but they do not enter the outer-halo region. \citet{2009ApJ...702.1058Z} suggested that in about 70\%-90\% of halo stars, for Milky Way-like galaxies, have an extra-galactic origin. In addition, in situ stars reside in the halo's inner few kpc, accreted stars dominate the stellar halo at all radii. \citet{2013ApJ...763...65A} constructed an in situ metallicity distribution function (MDF) of the Galactic halo, where a single Gaussian, with peak at [Fe/H] = $-1.5$, can adequately fit this MDF. Similar results have been reported by \citet{2017ApJ...845..101B}, who suggested that metal-rich halo stars in the solar neighborhood actually formed in situ within the Galactic disk, rather than having been accreted from satellite systems.

Many authors have observed and discussed the CEMP-no abundance signature among the classical dwarf spheroidal galaxies (dSphs) and ultra-faint dwarf galaxies (UFDs) \citep[see for example,][]{2015ARA&A..53..631F,2015MNRAS.454.1320S,2017MNRAS.465.2212S,2018A&A...617A..56S,2019ApJ...878...97Y,2020ApJ...894....7Y}. The Massive dSphs ($M/M_{\odot} \approx 10^9$) had much higher gas content than the UFDs ($M/M_{\odot} \approx 10^6$) at the time of their formation; given their much larger masses \citep{2008MNRAS.386..348S}. This mass range is resulting that dSphs have more extensive star formation events. While UFDs experience much more metal-poor star formation environments \citep[e.g., ][]{2012ApJ...759..115F}. The UFDs stars are predominantly found at extremely low metallicity, where most halo CEMP-no stars are also found.

Collectively, the low metallicity nature ([Fe/H] $< -2.9$), carbon enhancement, and kinematics of our sample indicate that these stars might have not been formed in-situ, but are most likely formed in now-destroyed satellites and accreted later on. Therefore, a comparison with the newly discovered accretion events the $Gaia$-Sausage \citep{2018MNRAS.478..611B,2018ApJ...863..113H,2018Natur.563...85H} and the $Gaia$-Sequoia event \citep{2019MNRAS.488.1235M} might help us to investigate their accretion possibility.

\begin{figure*}[!hptb]
\centerline{\includegraphics[width=600pt,height=30pc]{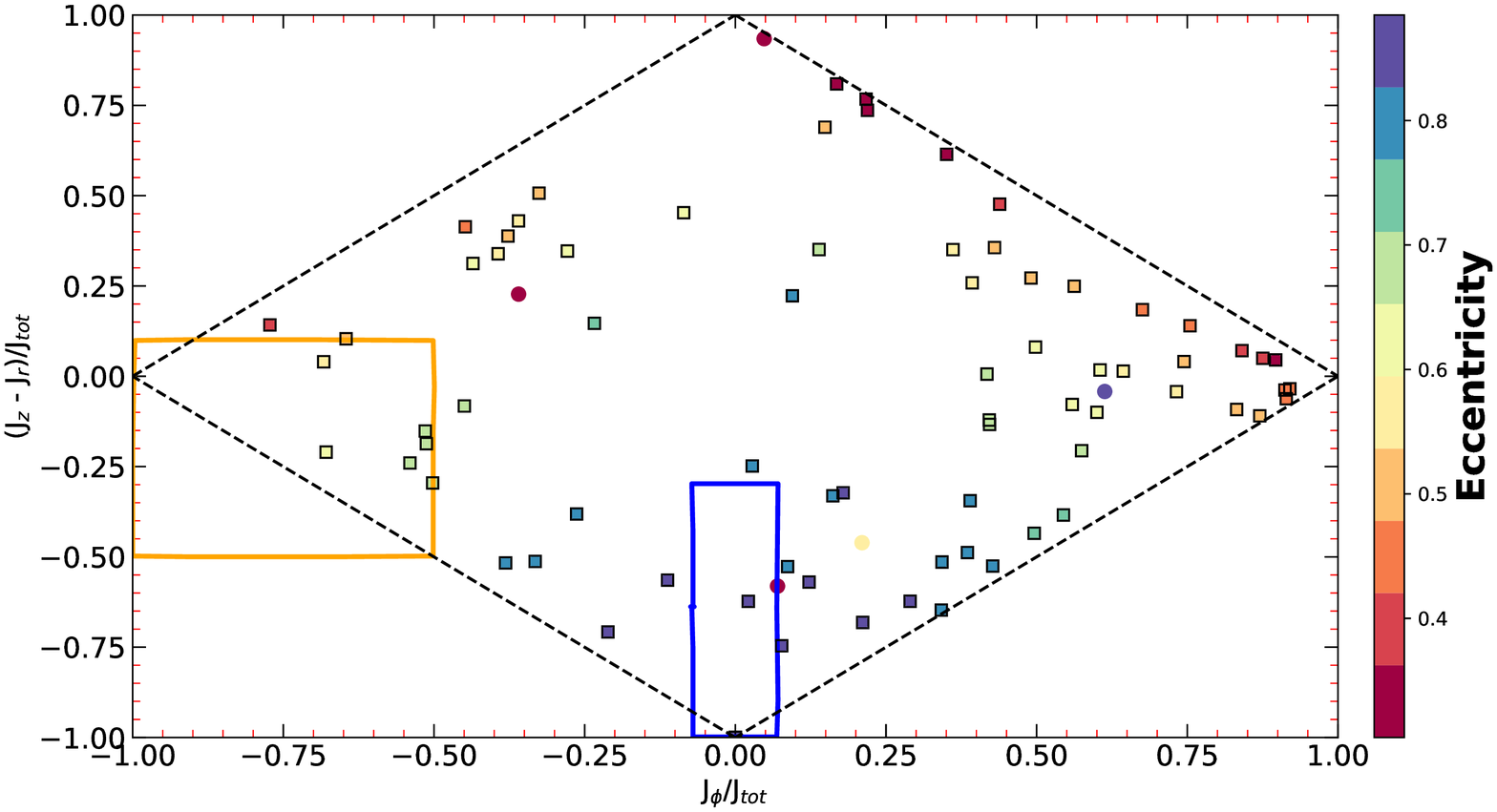}}
\caption{The action-space-map of our sample colored by their eccentricities. The orange and blue boxes denote the approximate locations of $Gaia$-Sequoia and $Gaia$-Enceladus-Sausage, respectively. The black points represent other literature CEMP-no adopted from JINAbase \citep{2018ApJS..238...36A}.\label{fig:Fig3}}
\end{figure*}

Figure~\ref{fig:Fig3} shows the action-space map for our sample and some literature CEMP-no stars adopted from JINAbase \citep{2018ApJS..238...36A}, overlaid with the approximate position of the $Gaia$-Sausage (blue box) and the $Gaia$-Sequoia event (orange box). Filled circles denote our sample stars, while filled squares represent other literature CEMP-no stars. The peak of the metallicity-distribution of $Gaia$-Sausage ([Fe/H] $\sim -1.3$) and the $Gaia$-Sequoia ([Fe/H] $\sim -1.6$) events suggest that all the CEMP-no are most likely not belonging to these accretion events. Numerically, eccentric stars (e $\sim$ 0.9) with $|J_\phi /J_{tot}|$ $< 0.07$ and (J$_Z$ - J$_R$)/$J_{tot}$ $< -0.3$ are most likely the $Gaia$-Sausage remnant stars. While less eccentric stars (e $\sim$ 0.6) with J$_\phi$/J$_{tot}$ $< -0.5$ and (J$_Z$ $-$ J$_R$)/J$_{tot}$ $< 0.1$ are most likely the $Gaia$-Sequoia remnant stars
\citep{2019MNRAS.488.1235M}. Using these values, three stars might be part of the $Gaia$-Sausage and seven stars might being part of the $Gaia$-Sequoia. However, the extremely low metallicity nature of these stars ([Fe/H]$ < -3.0$) refrains us from making such a conclusion. While the remaining CEMP-no stars are unambiguously not belonging to these accretion events. This suggests that if the CEMP-no stars have an extra-galactic origin, another accretion event should be responsible for adding of CEMP-no stars to the Galactic halo. More interestingly, J1630+0953 has a more circular orbit than the rest of our sample, but it shows a high variance between its J$_z$ and J$_r$. This might be unique dynamical properties. However, the small sample size precluding any solid opinion.

\section{Conclusions} \label{sec:concs}

This study carried out a statistical analysis of the progenitors' possible stellar masses, the phase space coordinates, and orbital parameters to investigate the chemo-dynamical evolution of five CEMP-no stars. The direct comparison between the chemical abundances and the predicted SN yields, suggests possible progenitors of our CEMP-no stars, with stellar-mass span the range of 11-22\,M$_{\odot}$ and explosion energies range of $0.3-1.8 \times 10^{51}$ erg. The chemical patterns and kinematic signatures of the five CEMP-no stars, suggest an extra-galactic origin such as a low-mass dwarf galaxy. In general, this chemo-dynamical analysis will provide information on the chemical evolution and the assembly history of the Galactic halo.
 
The action-space map of these CEMP-no stars seems not to favor the $Gaia$-Sausage and $Gaia$-Sequoia memberships, instead of that, it suggests that another minor accretion event might be responsible for the contribution of these CEMP-no stars to the Milky-Way. Future CEMP-no stars observations will provide crucial information to investigate their accretion origin.

\color{black}
\section*{acknowledgements}

We thank Elisabetta Caffau for perceptive and constructive comments, which led to significant improvements in the manuscript. This work has made use of data from the European Space Agency (ESA) mission
{\it Gaia} (\url{https://www.cosmos.esa.int/gaia}), processed by the {\it Gaia}
Data Processing and Analysis Consortium (DPAC, \url{https://www.cosmos.esa.int/web/gaia/dpac/consortium}). Funding for the DPAC
has been provided by national institutions, in particular the institutions
participating in the {\it Gaia} Multilateral Agreement.

\bibliography{CEMP}%

\begin{thebibliography}{}

\bibitem [\protect \citeauthoryear {%
{Abohalima}%
\ \BBA {} {Frebel}%
}{%
{Abohalima}%
\ \BBA {} {Frebel}%
}{%
{\protect \APACyear {2018}}%
}]{%
2018ApJS..238...36A}
\APACinsertmetastar {%
2018ApJS..238...36A}%
\begin{APACrefauthors}%
{Abohalima}, A.%
\BCBT {}\ \BBA {} {Frebel}, A.%
\end{APACrefauthors}%
\unskip\
\newblock
\APACrefYearMonthDay{2018}{Oct}{},
\newblock
\unskip
\newblock
\APACjournalVolNumPages{\apjs}{238}{2}{36}.
\newblock
\begin{APACrefDOI} \doi{10.3847/1538-4365/aadfe9} \end{APACrefDOI}
\PrintBackRefs{\CurrentBib}

\bibitem [\protect \citeauthoryear {%
{An}%
\ \protect \BOthers {.}}{%
{An}%
\ \protect \BOthers {.}}{%
{\protect \APACyear {2013}}%
}]{%
2013ApJ...763...65A}
\APACinsertmetastar {%
2013ApJ...763...65A}%
\begin{APACrefauthors}%
{An}, D.%
, {Beers}, T\BPBI C.%
, {Johnson}, J\BPBI A.%
\ et al.\end{APACrefauthors}%
\unskip\
\newblock
\APACrefYearMonthDay{2013}{{\APACmonth{01}}}{},
\newblock
\unskip
\newblock
\APACjournalVolNumPages{\apj}{763}{1}{65}.
\newblock
\begin{APACrefDOI} \doi{10.1088/0004-637X/763/1/65} \end{APACrefDOI}
\PrintBackRefs{\CurrentBib}

\bibitem [\protect \citeauthoryear {%
{Aoki}%
\ \protect \BOthers {.}}{%
{Aoki}%
\ \protect \BOthers {.}}{%
{\protect \APACyear {2007}}%
}]{%
2007ApJ...655..492A}
\APACinsertmetastar {%
2007ApJ...655..492A}%
\begin{APACrefauthors}%
{Aoki}, W.%
, {Beers}, T\BPBI C.%
, {Christlieb}, N.%
, {Norris}, J\BPBI E.%
, {Ryan}, S\BPBI G.%
\BCBL {}\ \BBA {} {Tsangarides}, S.%
\end{APACrefauthors}%
\unskip\
\newblock
\APACrefYearMonthDay{2007}{{\APACmonth{01}}}{},
\newblock
\unskip
\newblock
\APACjournalVolNumPages{\apj}{655}{1}{492-521}.
\newblock
\begin{APACrefDOI} \doi{10.1086/509817} \end{APACrefDOI}
\PrintBackRefs{\CurrentBib}

\bibitem [\protect \citeauthoryear {%
{Bailer-Jones et al.}%
}{%
{Bailer-Jones et al.}%
}{%
{\protect \APACyear {2018}}%
}]{%
BJones18}
\APACinsertmetastar {%
BJones18}%
\begin{APACrefauthors}%
{Bailer-Jones et al.}%
\end{APACrefauthors}%
\unskip\
\newblock
\APACrefYearMonthDay{2018}{{\APACmonth{08}}}{},
\newblock
\unskip
\newblock
\APACjournalVolNumPages{\aj}{156}{}{58}.
\newblock
\begin{APACrefDOI} \doi{10.3847/1538-3881/aacb21} \end{APACrefDOI}
\PrintBackRefs{\CurrentBib}

\bibitem [\protect \citeauthoryear {%
{Beers}%
\ \BBA {} {Christlieb}%
}{%
{Beers}%
\ \BBA {} {Christlieb}%
}{%
{\protect \APACyear {2005}}%
}]{%
2005ARA&A..43..531B}
\APACinsertmetastar {%
2005ARA&A..43..531B}%
\begin{APACrefauthors}%
{Beers}, T\BPBI C.%
\BCBT {}\ \BBA {} {Christlieb}, N.%
\end{APACrefauthors}%
\unskip\
\newblock
\APACrefYearMonthDay{2005}{{\APACmonth{09}}}{},
\newblock
\unskip
\newblock
\APACjournalVolNumPages{\araa}{43}{1}{531-580}.
\newblock
\begin{APACrefDOI} \doi{10.1146/annurev.astro.42.053102.134057}
  \end{APACrefDOI}
\PrintBackRefs{\CurrentBib}

\bibitem [\protect \citeauthoryear {%
{Belokurov}%
, {Erkal}%
, {Evans}%
, {Koposov}%
\BCBL {}\ \BBA {} {Deason}%
}{%
{Belokurov}%
\ \protect \BOthers {.}}{%
{\protect \APACyear {2018}}%
}]{%
2018MNRAS.478..611B}
\APACinsertmetastar {%
2018MNRAS.478..611B}%
\begin{APACrefauthors}%
{Belokurov}, V.%
, {Erkal}, D.%
, {Evans}, N\BPBI W.%
, {Koposov}, S\BPBI E.%
\BCBL {}\ \BBA {} {Deason}, A\BPBI J.%
\end{APACrefauthors}%
\unskip\
\newblock
\APACrefYearMonthDay{2018}{Jul}{},
\newblock
\unskip
\newblock
\APACjournalVolNumPages{\mnras}{478}{1}{611-619}.
\newblock
\begin{APACrefDOI} \doi{10.1093/mnras/sty982} \end{APACrefDOI}
\PrintBackRefs{\CurrentBib}

\bibitem [\protect \citeauthoryear {%
{Bennett}%
\ \BBA {} {Bovy}%
}{%
{Bennett}%
\ \BBA {} {Bovy}%
}{%
{\protect \APACyear {2019}}%
}]{%
2019MNRAS.482.1417B}
\APACinsertmetastar {%
2019MNRAS.482.1417B}%
\begin{APACrefauthors}%
{Bennett}, M.%
\BCBT {}\ \BBA {} {Bovy}, J.%
\end{APACrefauthors}%
\unskip\
\newblock
\APACrefYearMonthDay{2019}{{\APACmonth{01}}}{},
\newblock
\unskip
\newblock
\APACjournalVolNumPages{\mnras}{482}{1}{1417-1425}.
\newblock
\begin{APACrefDOI} \doi{10.1093/mnras/sty2813} \end{APACrefDOI}
\PrintBackRefs{\CurrentBib}

\bibitem [\protect \citeauthoryear {%
{Bonaca}%
, {Conroy}%
, {Wetzel}%
, {Hopkins}%
\BCBL {}\ \BBA {} {Kere{\v{s}}}%
}{%
{Bonaca}%
\ \protect \BOthers {.}}{%
{\protect \APACyear {2017}}%
}]{%
2017ApJ...845..101B}
\APACinsertmetastar {%
2017ApJ...845..101B}%
\begin{APACrefauthors}%
{Bonaca}, A.%
, {Conroy}, C.%
, {Wetzel}, A.%
, {Hopkins}, P\BPBI F.%
\BCBL {}\ \BBA {} {Kere{\v{s}}}, D.%
\end{APACrefauthors}%
\unskip\
\newblock
\APACrefYearMonthDay{2017}{{\APACmonth{08}}}{},
\newblock
\unskip
\newblock
\APACjournalVolNumPages{\apj}{845}{2}{101}.
\newblock
\begin{APACrefDOI} \doi{10.3847/1538-4357/aa7d0c} \end{APACrefDOI}
\PrintBackRefs{\CurrentBib}

\bibitem [\protect \citeauthoryear {%
{Bonifacio}%
\ \protect \BOthers {.}}{%
{Bonifacio}%
\ \protect \BOthers {.}}{%
{\protect \APACyear {2009}}%
}]{%
2009A&A...501..519B}
\APACinsertmetastar {%
2009A&A...501..519B}%
\begin{APACrefauthors}%
{Bonifacio}, P.%
, {Spite}, M.%
, {Cayrel}, R.%
\ et al.\end{APACrefauthors}%
\unskip\
\newblock
\APACrefYearMonthDay{2009}{{\APACmonth{07}}}{},
\newblock
\unskip
\newblock
\APACjournalVolNumPages{\aap}{501}{2}{519-530}.
\newblock
\begin{APACrefDOI} \doi{10.1051/0004-6361/200810610} \end{APACrefDOI}
\PrintBackRefs{\CurrentBib}

\bibitem [\protect \citeauthoryear {%
{Bovy}%
}{%
{Bovy}%
}{%
{\protect \APACyear {2015}}%
}]{%
Bovy15}
\APACinsertmetastar {%
Bovy15}%
\begin{APACrefauthors}%
{Bovy}, J.%
\end{APACrefauthors}%
\unskip\
\newblock
\APACrefYearMonthDay{2015}{{\APACmonth{02}}}{},
\newblock
\unskip
\newblock
\APACjournalVolNumPages{\apjs}{216}{2}{29}.
\newblock
\begin{APACrefDOI} \doi{10.1088/0067-0049/216/2/29} \end{APACrefDOI}
\PrintBackRefs{\CurrentBib}

\bibitem [\protect \citeauthoryear {%
{Burris}%
\ \protect \BOthers {.}}{%
{Burris}%
\ \protect \BOthers {.}}{%
{\protect \APACyear {2000}}%
}]{%
2000ApJ...544..302B}
\APACinsertmetastar {%
2000ApJ...544..302B}%
\begin{APACrefauthors}%
{Burris}, D\BPBI L.%
, {Pilachowski}, C\BPBI A.%
, {Armand roff}, T\BPBI E.%
, {Sneden}, C.%
, {Cowan}, J\BPBI J.%
\BCBL {}\ \BBA {} {Roe}, H.%
\end{APACrefauthors}%
\unskip\
\newblock
\APACrefYearMonthDay{2000}{{\APACmonth{11}}}{},
\newblock
\unskip
\newblock
\APACjournalVolNumPages{\apj}{544}{1}{302-319}.
\newblock
\begin{APACrefDOI} \doi{10.1086/317172} \end{APACrefDOI}
\PrintBackRefs{\CurrentBib}

\bibitem [\protect \citeauthoryear {%
{Caffau}%
\ \protect \BOthers {.}}{%
{Caffau}%
\ \protect \BOthers {.}}{%
{\protect \APACyear {2011}}%
}]{%
2011Natur.477...67C}
\APACinsertmetastar {%
2011Natur.477...67C}%
\begin{APACrefauthors}%
{Caffau}, E.%
, {Bonifacio}, P.%
, {Fran{\c{c}}ois}, P.%
\ et al.\end{APACrefauthors}%
\unskip\
\newblock
\APACrefYearMonthDay{2011}{{\APACmonth{09}}}{},
\newblock
\unskip
\newblock
\APACjournalVolNumPages{\nat}{477}{7362}{67-69}.
\newblock
\begin{APACrefDOI} \doi{10.1038/nature10377} \end{APACrefDOI}
\PrintBackRefs{\CurrentBib}

\bibitem [\protect \citeauthoryear {%
{Cayrel}%
\ \protect \BOthers {.}}{%
{Cayrel}%
\ \protect \BOthers {.}}{%
{\protect \APACyear {2004}}%
}]{%
2004A&A...416.1117C}
\APACinsertmetastar {%
2004A&A...416.1117C}%
\begin{APACrefauthors}%
{Cayrel}, R.%
, {Depagne}, E.%
, {Spite}, M.%
\ et al.\end{APACrefauthors}%
\unskip\
\newblock
\APACrefYearMonthDay{2004}{{\APACmonth{03}}}{},
\newblock
\unskip
\newblock
\APACjournalVolNumPages{\aap}{416}{}{1117-1138}.
\newblock
\begin{APACrefDOI} \doi{10.1051/0004-6361:20034074} \end{APACrefDOI}
\PrintBackRefs{\CurrentBib}

\bibitem [\protect \citeauthoryear {%
{Chiti}%
, {Frebel}%
\BCBL {}\ \protect \BOthers {.}}{%
{Chiti}%
, {Frebel}%
\BCBL {}\ \protect \BOthers {.}}{%
{\protect \APACyear {2021}}%
}]{%
2021arXiv210316660C}
\APACinsertmetastar {%
2021arXiv210316660C}%
\begin{APACrefauthors}%
{Chiti}, A.%
, {Frebel}, A.%
, {Mardini}, M\BPBI K.%
, {Daniel}, T\BPBI W.%
, {Ou}, X.%
\BCBL {}\ \BBA {} {Uvarova}, A\BPBI V.%
\end{APACrefauthors}%
\unskip\
\newblock
\APACrefYearMonthDay{2021}{{\APACmonth{03}}}{},
\newblock
\unskip
\newblock
\APACjournalVolNumPages{arXiv e-prints}{}{}{arXiv:2103.16660}.
\PrintBackRefs{\CurrentBib}

\bibitem [\protect \citeauthoryear {%
{Chiti}%
, {Mardini}%
, {Frebel}%
\BCBL {}\ \BBA {} {Daniel}%
}{%
{Chiti}%
, {Mardini}%
\BCBL {}\ \protect \BOthers {.}}{%
{\protect \APACyear {2021}}%
}]{%
2021arXiv210316642C}
\APACinsertmetastar {%
2021arXiv210316642C}%
\begin{APACrefauthors}%
{Chiti}, A.%
, {Mardini}, M\BPBI K.%
, {Frebel}, A.%
\BCBL {}\ \BBA {} {Daniel}, T.%
\end{APACrefauthors}%
\unskip\
\newblock
\APACrefYearMonthDay{2021}{{\APACmonth{03}}}{},
\newblock
\unskip
\newblock
\APACjournalVolNumPages{arXiv e-prints}{}{}{arXiv:2103.16642}.
\PrintBackRefs{\CurrentBib}

\bibitem [\protect \citeauthoryear {%
{Cui}%
, {Zhao}%
\BCBL {}\ \BBA {} {Chu}%
}{%
{Cui}%
\ \protect \BOthers {.}}{%
{\protect \APACyear {2012}}%
}]{%
2012RAA....12.1197C}
\APACinsertmetastar {%
2012RAA....12.1197C}%
\begin{APACrefauthors}%
{Cui}, X\BHBI Q.%
, {Zhao}, Y\BHBI H.%
\BCBL {}\ \BBA {} {Chu}, Y\BHBI Q.%
\end{APACrefauthors}%
\unskip\
\newblock
\APACrefYearMonthDay{2012}{{\APACmonth{09}}}{},
\newblock
\unskip
\newblock
\APACjournalVolNumPages{Research in Astronomy and
  Astrophysics}{12}{9}{1197-1242}.
\newblock
\begin{APACrefDOI} \doi{10.1088/1674-4527/12/9/003} \end{APACrefDOI}
\PrintBackRefs{\CurrentBib}

\bibitem [\protect \citeauthoryear {%
{Di Matteo}%
\ \protect \BOthers {.}}{%
{Di Matteo}%
\ \protect \BOthers {.}}{%
{\protect \APACyear {2020}}%
}]{%
2020A&A...636A.115D}
\APACinsertmetastar {%
2020A&A...636A.115D}%
\begin{APACrefauthors}%
{Di Matteo}, P.%
, {Spite}, M.%
, {Haywood}, M.%
, {Bonifacio}, P.%
, {G{\'o}mez}, A.%
, {Spite}, F.%
\BCBL {}\ \BBA {} {Caffau}, E.%
\end{APACrefauthors}%
\unskip\
\newblock
\APACrefYearMonthDay{2020}{{\APACmonth{04}}}{},
\newblock
\unskip
\newblock
\APACjournalVolNumPages{\aap}{636}{}{A115}.
\newblock
\begin{APACrefDOI} \doi{10.1051/0004-6361/201937016} \end{APACrefDOI}
\PrintBackRefs{\CurrentBib}

\bibitem [\protect \citeauthoryear {%
{Frebel}%
\ \BBA {} {Bromm}%
}{%
{Frebel}%
\ \BBA {} {Bromm}%
}{%
{\protect \APACyear {2012}}%
}]{%
2012ApJ...759..115F}
\APACinsertmetastar {%
2012ApJ...759..115F}%
\begin{APACrefauthors}%
{Frebel}, A.%
\BCBT {}\ \BBA {} {Bromm}, V.%
\end{APACrefauthors}%
\unskip\
\newblock
\APACrefYearMonthDay{2012}{{\APACmonth{11}}}{},
\newblock
\unskip
\newblock
\APACjournalVolNumPages{\apj}{759}{2}{115}.
\newblock
\begin{APACrefDOI} \doi{10.1088/0004-637X/759/2/115} \end{APACrefDOI}
\PrintBackRefs{\CurrentBib}

\bibitem [\protect \citeauthoryear {%
{Frebel}%
\ \BBA {} {Norris}%
}{%
{Frebel}%
\ \BBA {} {Norris}%
}{%
{\protect \APACyear {2015}}%
}]{%
2015ARA&A..53..631F}
\APACinsertmetastar {%
2015ARA&A..53..631F}%
\begin{APACrefauthors}%
{Frebel}, A.%
\BCBT {}\ \BBA {} {Norris}, J\BPBI E.%
\end{APACrefauthors}%
\unskip\
\newblock
\APACrefYearMonthDay{2015}{{\APACmonth{08}}}{},
\newblock
\unskip
\newblock
\APACjournalVolNumPages{\araa}{53}{}{631-688}.
\newblock
\begin{APACrefDOI} \doi{10.1146/annurev-astro-082214-122423} \end{APACrefDOI}
\PrintBackRefs{\CurrentBib}

\bibitem [\protect \citeauthoryear {%
{Frebel}%
\ \BBA {} {Surman}%
}{%
{Frebel}%
\ \BBA {} {Surman}%
}{%
{\protect \APACyear {2020}}%
}]{%
2020NatAs...4..564F}
\APACinsertmetastar {%
2020NatAs...4..564F}%
\begin{APACrefauthors}%
{Frebel}, A.%
\BCBT {}\ \BBA {} {Surman}, R.%
\end{APACrefauthors}%
\unskip\
\newblock
\APACrefYearMonthDay{2020}{{\APACmonth{05}}}{},
\newblock
\unskip
\newblock
\APACjournalVolNumPages{Nature Astronomy}{4}{}{564-565}.
\newblock
\begin{APACrefDOI} \doi{10.1038/s41550-020-1115-2} \end{APACrefDOI}
\PrintBackRefs{\CurrentBib}

\bibitem [\protect \citeauthoryear {%
{Gaia Collaboration}%
\ \protect \BOthers {.}}{%
{Gaia Collaboration}%
\ \protect \BOthers {.}}{%
{\protect \APACyear {2018}}%
}]{%
2018A&A...616A...1G}
\APACinsertmetastar {%
2018A&A...616A...1G}%
\begin{APACrefauthors}%
{Gaia Collaboration}%
, {Brown}, A\BPBI G\BPBI A.%
, {Vallenari}, A.%
\ et al.\end{APACrefauthors}%
\unskip\
\newblock
\APACrefYearMonthDay{2018}{Aug}{},
\newblock
\unskip
\newblock
\APACjournalVolNumPages{\aap}{616}{}{A1}.
\newblock
\begin{APACrefDOI} \doi{10.1051/0004-6361/201833051} \end{APACrefDOI}
\PrintBackRefs{\CurrentBib}

\bibitem [\protect \citeauthoryear {%
{Gravity Collaboration}%
, {Abuter}%
, {Amorim}%
\BCBL {}\ \BBA {} {Baub{\"o}ck}%
}{%
{Gravity Collaboration}%
\ \protect \BOthers {.}}{%
{\protect \APACyear {2019}}%
}]{%
2019A&A...625L..10G}
\APACinsertmetastar {%
2019A&A...625L..10G}%
\begin{APACrefauthors}%
{Gravity Collaboration}%
, {Abuter}, R.%
, {Amorim}, A.%
\BCBL {}\ \BBA {} {Baub{\"o}ck}, e\BPBI a.%
\end{APACrefauthors}%
\unskip\
\newblock
\APACrefYearMonthDay{2019}{{\APACmonth{05}}}{},
\newblock
\unskip
\newblock
\APACjournalVolNumPages{\aap}{625}{}{L10}.
\newblock
\begin{APACrefDOI} \doi{10.1051/0004-6361/201935656} \end{APACrefDOI}
\PrintBackRefs{\CurrentBib}

\bibitem [\protect \citeauthoryear {%
{Haywood}%
\ \protect \BOthers {.}}{%
{Haywood}%
\ \protect \BOthers {.}}{%
{\protect \APACyear {2018}}%
}]{%
2018ApJ...863..113H}
\APACinsertmetastar {%
2018ApJ...863..113H}%
\begin{APACrefauthors}%
{Haywood}, M.%
, {Di Matteo}, P.%
, {Lehnert}, M\BPBI D.%
, {Snaith}, O.%
, {Khoperskov}, S.%
\BCBL {}\ \BBA {} {G{\'o}mez}, A.%
\end{APACrefauthors}%
\unskip\
\newblock
\APACrefYearMonthDay{2018}{Aug}{},
\newblock
\unskip
\newblock
\APACjournalVolNumPages{\apj}{863}{2}{113}.
\newblock
\begin{APACrefDOI} \doi{10.3847/1538-4357/aad235} \end{APACrefDOI}
\PrintBackRefs{\CurrentBib}

\bibitem [\protect \citeauthoryear {%
{Heger}%
\ \BBA {} {Woosley}%
}{%
{Heger}%
\ \BBA {} {Woosley}%
}{%
{\protect \APACyear {2010}}%
}]{%
Hegerwoosley10}
\APACinsertmetastar {%
Hegerwoosley10}%
\begin{APACrefauthors}%
{Heger}, A.%
\BCBT {}\ \BBA {} {Woosley}, S\BPBI E.%
\end{APACrefauthors}%
\unskip\
\newblock
\APACrefYearMonthDay{2010}{{\APACmonth{11}}}{},
\newblock
\unskip
\newblock
\APACjournalVolNumPages{\apj}{724}{1}{341-373}.
\newblock
\begin{APACrefDOI} \doi{10.1088/0004-637X/724/1/341} \end{APACrefDOI}
\PrintBackRefs{\CurrentBib}

\bibitem [\protect \citeauthoryear {%
{Helmi}%
, {Babusiaux}%
\BCBL {}\ \BBA {} {Koppelman}%
}{%
{Helmi}%
\ \protect \BOthers {.}}{%
{\protect \APACyear {2018}}%
}]{%
2018Natur.563...85H}
\APACinsertmetastar {%
2018Natur.563...85H}%
\begin{APACrefauthors}%
{Helmi}, A.%
, {Babusiaux}, C.%
\BCBL {}\ \BBA {} {Koppelman}, H\BPBI H\BPBI e\BPBI a.%
\end{APACrefauthors}%
\unskip\
\newblock
\APACrefYearMonthDay{2018}{Nov}{},
\newblock
\unskip
\newblock
\APACjournalVolNumPages{\nat}{563}{7729}{85-88}.
\newblock
\begin{APACrefDOI} \doi{10.1038/s41586-018-0625-x} \end{APACrefDOI}
\PrintBackRefs{\CurrentBib}

\bibitem [\protect \citeauthoryear {%
{Hutchins}%
}{%
{Hutchins}%
}{%
{\protect \APACyear {1976}}%
}]{%
1976ApJ...205..103H}
\APACinsertmetastar {%
1976ApJ...205..103H}%
\begin{APACrefauthors}%
{Hutchins}, J\BPBI B.%
\end{APACrefauthors}%
\unskip\
\newblock
\APACrefYearMonthDay{1976}{{\APACmonth{04}}}{},
\newblock
\unskip
\newblock
\APACjournalVolNumPages{\apj}{205}{}{103-121}.
\newblock
\begin{APACrefDOI} \doi{10.1086/154254} \end{APACrefDOI}
\PrintBackRefs{\CurrentBib}

\bibitem [\protect \citeauthoryear {%
{Ishigaki et al.}%
}{%
{Ishigaki et al.}%
}{%
{\protect \APACyear {2018}}%
}]{%
Ishigaki18}
\APACinsertmetastar {%
Ishigaki18}%
\begin{APACrefauthors}%
{Ishigaki et al.}%
\end{APACrefauthors}%
\unskip\
\newblock
\APACrefYearMonthDay{2018}{{\APACmonth{04}}}{},
\newblock
\unskip
\newblock
\APACjournalVolNumPages{\apj}{857}{1}{46}.
\newblock
\begin{APACrefDOI} \doi{10.3847/1538-4357/aab3de} \end{APACrefDOI}
\PrintBackRefs{\CurrentBib}

\bibitem [\protect \citeauthoryear {%
{Lindegren}%
, {Hern{\'a}ndez}%
\BCBL {}\ \BBA {} {Bombrun}%
}{%
{Lindegren}%
\ \protect \BOthers {.}}{%
{\protect \APACyear {2018}}%
}]{%
2018A&A...616A...2L}
\APACinsertmetastar {%
2018A&A...616A...2L}%
\begin{APACrefauthors}%
{Lindegren}, L.%
, {Hern{\'a}ndez}, J.%
\BCBL {}\ \BBA {} {Bombrun}.%
\end{APACrefauthors}%
\unskip\
\newblock
\APACrefYearMonthDay{2018}{{\APACmonth{08}}}{},
\newblock
\unskip
\newblock
\APACjournalVolNumPages{\aap}{616}{}{A2}.
\newblock
\begin{APACrefDOI} \doi{10.1051/0004-6361/201832727} \end{APACrefDOI}
\PrintBackRefs{\CurrentBib}

\bibitem [\protect \citeauthoryear {%
Mardini%
, Li%
\BCBL {}\ \protect \BOthers {.}}{%
Mardini%
, Li%
\BCBL {}\ \protect \BOthers {.}}{%
{\protect \APACyear {2019}}%
}]{%
Mardini_2019a}
\APACinsertmetastar {%
Mardini_2019a}%
\begin{APACrefauthors}%
Mardini, M\BPBI K.%
, Li, H.%
, Placco, V\BPBI M.%
\ et al.\end{APACrefauthors}%
\unskip\
\newblock
\APACrefYearMonthDay{2019}{apr}{},
\newblock
\unskip
\newblock
\APACjournalVolNumPages{The Astrophysical Journal}{875}{2}{89}.
\newblock
\begin{APACrefURL} \url{https://doi.org/10.3847/1538-4357/ab0fa2}
  \end{APACrefURL}
\newblock
\begin{APACrefDOI} \doi{10.3847/1538-4357/ab0fa2} \end{APACrefDOI}
\PrintBackRefs{\CurrentBib}

\bibitem [\protect \citeauthoryear {%
Mardini%
\ \protect \BOthers {.}}{%
Mardini%
\ \protect \BOthers {.}}{%
{\protect \APACyear {2020}}%
}]{%
Mardini_2020}
\APACinsertmetastar {%
Mardini_2020}%
\begin{APACrefauthors}%
Mardini, M\BPBI K.%
, Placco, V\BPBI M.%
, Meiron, Y.%
\ et al.\end{APACrefauthors}%
\unskip\
\newblock
\APACrefYearMonthDay{2020}{nov}{},
\newblock
\unskip
\newblock
\APACjournalVolNumPages{The Astrophysical Journal}{903}{2}{88}.
\newblock
\begin{APACrefURL} \url{https://doi.org/10.3847/1538-4357/abbc13}
  \end{APACrefURL}
\newblock
\begin{APACrefDOI} \doi{10.3847/1538-4357/abbc13} \end{APACrefDOI}
\PrintBackRefs{\CurrentBib}

\bibitem [\protect \citeauthoryear {%
Mardini%
, Placco%
, Taani%
, Li%
\BCBL {}\ \BBA {} Zhao%
}{%
Mardini%
, Placco%
\BCBL {}\ \protect \BOthers {.}}{%
{\protect \APACyear {2019}}%
}]{%
Mardini_2019b}
\APACinsertmetastar {%
Mardini_2019b}%
\begin{APACrefauthors}%
Mardini, M\BPBI K.%
, Placco, V\BPBI M.%
, Taani, A.%
, Li, H.%
\BCBL {}\ \BBA {} Zhao, G.%
\end{APACrefauthors}%
\unskip\
\newblock
\APACrefYearMonthDay{2019}{aug}{},
\newblock
\unskip
\newblock
\APACjournalVolNumPages{The Astrophysical Journal}{882}{1}{27}.
\newblock
\begin{APACrefURL} \url{https://doi.org/10.3847/1538-4357/ab3047}
  \end{APACrefURL}
\newblock
\begin{APACrefDOI} \doi{10.3847/1538-4357/ab3047} \end{APACrefDOI}
\PrintBackRefs{\CurrentBib}

\bibitem [\protect \citeauthoryear {%
{Masda}%
\ \protect \BOthers {.}}{%
{Masda}%
\ \protect \BOthers {.}}{%
{\protect \APACyear {2019}}%
}]{%
2019AstBu..74..464M}
\APACinsertmetastar {%
2019AstBu..74..464M}%
\begin{APACrefauthors}%
{Masda}, S\BPBI G.%
, {Docobo}, J\BPBI A.%
, {Hussein}, A\BPBI M.%
\ et al.\end{APACrefauthors}%
\unskip\
\newblock
\APACrefYearMonthDay{2019}{{\APACmonth{12}}}{},
\newblock
\unskip
\newblock
\APACjournalVolNumPages{Astrophysical Bulletin}{74}{4}{464-474}.
\newblock
\begin{APACrefDOI} \doi{10.1134/S1990341319040126} \end{APACrefDOI}
\PrintBackRefs{\CurrentBib}

\bibitem [\protect \citeauthoryear {%
{Matsuda}%
, {Sat{\={o}}}%
\BCBL {}\ \BBA {} {Takeda}%
}{%
{Matsuda}%
\ \protect \BOthers {.}}{%
{\protect \APACyear {1969}}%
}]{%
1969PThPh..42..219M}
\APACinsertmetastar {%
1969PThPh..42..219M}%
\begin{APACrefauthors}%
{Matsuda}, T.%
, {Sat{\={o}}}, H.%
\BCBL {}\ \BBA {} {Takeda}, H.%
\end{APACrefauthors}%
\unskip\
\newblock
\APACrefYearMonthDay{1969}{{\APACmonth{08}}}{},
\newblock
\unskip
\newblock
\APACjournalVolNumPages{Progress of Theoretical Physics}{42}{2}{219-233}.
\newblock
\begin{APACrefDOI} \doi{10.1143/PTP.42.219} \end{APACrefDOI}
\PrintBackRefs{\CurrentBib}

\bibitem [\protect \citeauthoryear {%
{Myeong}%
, {Vasiliev}%
, {Iorio}%
, {Evans}%
\BCBL {}\ \BBA {} {Belokurov}%
}{%
{Myeong}%
\ \protect \BOthers {.}}{%
{\protect \APACyear {2019}}%
}]{%
2019MNRAS.488.1235M}
\APACinsertmetastar {%
2019MNRAS.488.1235M}%
\begin{APACrefauthors}%
{Myeong}, G\BPBI C.%
, {Vasiliev}, E.%
, {Iorio}, G.%
, {Evans}, N\BPBI W.%
\BCBL {}\ \BBA {} {Belokurov}, V.%
\end{APACrefauthors}%
\unskip\
\newblock
\APACrefYearMonthDay{2019}{{\APACmonth{09}}}{},
\newblock
\unskip
\newblock
\APACjournalVolNumPages{\mnras}{488}{1}{1235-1247}.
\newblock
\begin{APACrefDOI} \doi{10.1093/mnras/stz1770} \end{APACrefDOI}
\PrintBackRefs{\CurrentBib}

\bibitem [\protect \citeauthoryear {%
{Placco}%
, {Frebel}%
, {Beers}%
\BCBL {}\ \BBA {} {Stancliffe}%
}{%
{Placco}%
\ \protect \BOthers {.}}{%
{\protect \APACyear {2014}}%
}]{%
2014ApJ...797...21P}
\APACinsertmetastar {%
2014ApJ...797...21P}%
\begin{APACrefauthors}%
{Placco}, V\BPBI M.%
, {Frebel}, A.%
, {Beers}, T\BPBI C.%
\BCBL {}\ \BBA {} {Stancliffe}, R\BPBI J.%
\end{APACrefauthors}%
\unskip\
\newblock
\APACrefYearMonthDay{2014}{{\APACmonth{12}}}{},
\newblock
\unskip
\newblock
\APACjournalVolNumPages{\apj}{797}{1}{21}.
\newblock
\begin{APACrefDOI} \doi{10.1088/0004-637X/797/1/21} \end{APACrefDOI}
\PrintBackRefs{\CurrentBib}

\bibitem [\protect \citeauthoryear {%
{Placco}%
\ \protect \BOthers {.}}{%
{Placco}%
\ \protect \BOthers {.}}{%
{\protect \APACyear {2016}}%
}]{%
2016ApJ...833...21P}
\APACinsertmetastar {%
2016ApJ...833...21P}%
\begin{APACrefauthors}%
{Placco}, V\BPBI M.%
, {Frebel}, A.%
, {Beers}, T\BPBI C.%
\ et al.\end{APACrefauthors}%
\unskip\
\newblock
\APACrefYearMonthDay{2016}{{\APACmonth{12}}}{},
\newblock
\unskip
\newblock
\APACjournalVolNumPages{\apj}{833}{1}{21}.
\newblock
\begin{APACrefDOI} \doi{10.3847/0004-637X/833/1/21} \end{APACrefDOI}
\PrintBackRefs{\CurrentBib}

\bibitem [\protect \citeauthoryear {%
{Placco}%
, {Santucci}%
\BCBL {}\ \BBA {} {Yuan}%
}{%
{Placco}%
\ \protect \BOthers {.}}{%
{\protect \APACyear {2020}}%
}]{%
2020ApJ...897...78P}
\APACinsertmetastar {%
2020ApJ...897...78P}%
\begin{APACrefauthors}%
{Placco}, V\BPBI M.%
, {Santucci}, R\BPBI M.%
\BCBL {}\ \BBA {} {Yuan}, Z.%
\end{APACrefauthors}%
\unskip\
\newblock
\APACrefYearMonthDay{2020}{{\APACmonth{07}}}{},
\newblock
\unskip
\newblock
\APACjournalVolNumPages{\apj}{897}{1}{78}.
\newblock
\begin{APACrefDOI} \doi{10.3847/1538-4357/ab99c6} \end{APACrefDOI}
\PrintBackRefs{\CurrentBib}

\bibitem [\protect \citeauthoryear {%
Price-Whelan%
, Hogg%
\BCBL {}\ \BBA {} et al.%
}{%
Price-Whelan%
\ \protect \BOthers {.}}{%
{\protect \APACyear {2018}}%
}]{%
Astropy18}
\APACinsertmetastar {%
Astropy18}%
\begin{APACrefauthors}%
Price-Whelan, A\BPBI M.%
, Hogg, D\BPBI W.%
\BCBL {}\ \BBA {} et al., H\BHBI W\BPBI R.%
\end{APACrefauthors}%
\unskip\
\newblock
\APACrefYearMonthDay{2018}{jun}{},
\newblock
\unskip
\newblock
\APACjournalVolNumPages{The Astronomical Journal}{156}{1}{18}.
\newblock
\begin{APACrefURL} \url{https://doi.org/10.3847%2F1538-3881%2Faac387}
  \end{APACrefURL}
\newblock
\begin{APACrefDOI} \doi{10.3847/1538-3881/aac387} \end{APACrefDOI}
\PrintBackRefs{\CurrentBib}

\bibitem [\protect \citeauthoryear {%
{Salvadori}%
, {Ferrara}%
\BCBL {}\ \BBA {} {Schneider}%
}{%
{Salvadori}%
\ \protect \BOthers {.}}{%
{\protect \APACyear {2008}}%
}]{%
2008MNRAS.386..348S}
\APACinsertmetastar {%
2008MNRAS.386..348S}%
\begin{APACrefauthors}%
{Salvadori}, S.%
, {Ferrara}, A.%
\BCBL {}\ \BBA {} {Schneider}, R.%
\end{APACrefauthors}%
\unskip\
\newblock
\APACrefYearMonthDay{2008}{{\APACmonth{05}}}{},
\newblock
\unskip
\newblock
\APACjournalVolNumPages{\mnras}{386}{1}{348-358}.
\newblock
\begin{APACrefDOI} \doi{10.1111/j.1365-2966.2008.13035.x} \end{APACrefDOI}
\PrintBackRefs{\CurrentBib}

\bibitem [\protect \citeauthoryear {%
{Salvadori}%
, {Sk{\'u}lad{\'o}ttir}%
\BCBL {}\ \BBA {} {Tolstoy}%
}{%
{Salvadori}%
\ \protect \BOthers {.}}{%
{\protect \APACyear {2015}}%
}]{%
2015MNRAS.454.1320S}
\APACinsertmetastar {%
2015MNRAS.454.1320S}%
\begin{APACrefauthors}%
{Salvadori}, S.%
, {Sk{\'u}lad{\'o}ttir}, {\'A}.%
\BCBL {}\ \BBA {} {Tolstoy}, E.%
\end{APACrefauthors}%
\unskip\
\newblock
\APACrefYearMonthDay{2015}{{\APACmonth{12}}}{},
\newblock
\unskip
\newblock
\APACjournalVolNumPages{\mnras}{454}{2}{1320-1331}.
\newblock
\begin{APACrefDOI} \doi{10.1093/mnras/stv1969} \end{APACrefDOI}
\PrintBackRefs{\CurrentBib}

\bibitem [\protect \citeauthoryear {%
{Sch{\"o}nrich}%
, {Binney}%
\BCBL {}\ \BBA {} {Dehnen}%
}{%
{Sch{\"o}nrich}%
\ \protect \BOthers {.}}{%
{\protect \APACyear {2010}}%
}]{%
2010MNRAS.403.1829S}
\APACinsertmetastar {%
2010MNRAS.403.1829S}%
\begin{APACrefauthors}%
{Sch{\"o}nrich}, R.%
, {Binney}, J.%
\BCBL {}\ \BBA {} {Dehnen}, W.%
\end{APACrefauthors}%
\unskip\
\newblock
\APACrefYearMonthDay{2010}{{\APACmonth{04}}}{},
\newblock
\unskip
\newblock
\APACjournalVolNumPages{\mnras}{403}{4}{1829-1833}.
\newblock
\begin{APACrefDOI} \doi{10.1111/j.1365-2966.2010.16253.x} \end{APACrefDOI}
\PrintBackRefs{\CurrentBib}

\bibitem [\protect \citeauthoryear {%
{Sestito}%
\ \protect \BOthers {.}}{%
{Sestito}%
\ \protect \BOthers {.}}{%
{\protect \APACyear {2019}}%
}]{%
2019MNRAS.484.2166S}
\APACinsertmetastar {%
2019MNRAS.484.2166S}%
\begin{APACrefauthors}%
{Sestito}, F.%
, {Longeard}, N.%
, {Martin}, N\BPBI F.%
\ et al.\end{APACrefauthors}%
\unskip\
\newblock
\APACrefYearMonthDay{2019}{{\APACmonth{04}}}{},
\newblock
\unskip
\newblock
\APACjournalVolNumPages{\mnras}{484}{2}{2166-2180}.
\newblock
\begin{APACrefDOI} \doi{10.1093/mnras/stz043} \end{APACrefDOI}
\PrintBackRefs{\CurrentBib}

\bibitem [\protect \citeauthoryear {%
{Silk}%
}{%
{Silk}%
}{%
{\protect \APACyear {1977}}%
{\protect \APACexlab {{\protect \BCnt {1}}}}}]{%
1977ApJ...214..152S}
\APACinsertmetastar {%
1977ApJ...214..152S}%
\begin{APACrefauthors}%
{Silk}, J.%
\end{APACrefauthors}%
\unskip\
\newblock
\APACrefYearMonthDay{1977{\protect \BCnt {1}}}{{\APACmonth{05}}}{},
\newblock
\unskip
\newblock
\APACjournalVolNumPages{\apj}{214}{}{152-160}.
\newblock
\begin{APACrefDOI} \doi{10.1086/155240} \end{APACrefDOI}
\PrintBackRefs{\CurrentBib}

\bibitem [\protect \citeauthoryear {%
{Silk}%
}{%
{Silk}%
}{%
{\protect \APACyear {1977}}%
{\protect \APACexlab {{\protect \BCnt {2}}}}}]{%
1977ApJ...211..638S}
\APACinsertmetastar {%
1977ApJ...211..638S}%
\begin{APACrefauthors}%
{Silk}, J.%
\end{APACrefauthors}%
\unskip\
\newblock
\APACrefYearMonthDay{1977{\protect \BCnt {2}}}{{\APACmonth{02}}}{},
\newblock
\unskip
\newblock
\APACjournalVolNumPages{\apj}{211}{}{638-648}.
\newblock
\begin{APACrefDOI} \doi{10.1086/154972} \end{APACrefDOI}
\PrintBackRefs{\CurrentBib}

\bibitem [\protect \citeauthoryear {%
{Spite}%
\ \protect \BOthers {.}}{%
{Spite}%
\ \protect \BOthers {.}}{%
{\protect \APACyear {2018}}%
}]{%
2018A&A...617A..56S}
\APACinsertmetastar {%
2018A&A...617A..56S}%
\begin{APACrefauthors}%
{Spite}, M.%
, {Spite}, F.%
, {Fran{\c{c}}ois}, P.%
, {Bonifacio}, P.%
, {Caffau}, E.%
\BCBL {}\ \BBA {} {Salvadori}, S.%
\end{APACrefauthors}%
\unskip\
\newblock
\APACrefYearMonthDay{2018}{{\APACmonth{09}}}{},
\newblock
\unskip
\newblock
\APACjournalVolNumPages{\aap}{617}{}{A56}.
\newblock
\begin{APACrefDOI} \doi{10.1051/0004-6361/201833548} \end{APACrefDOI}
\PrintBackRefs{\CurrentBib}

\bibitem [\protect \citeauthoryear {%
{Starkenburg}%
\ \protect \BOthers {.}}{%
{Starkenburg}%
\ \protect \BOthers {.}}{%
{\protect \APACyear {2018}}%
}]{%
2018MNRAS.481.3838S}
\APACinsertmetastar {%
2018MNRAS.481.3838S}%
\begin{APACrefauthors}%
{Starkenburg}, E.%
, {Aguado}, D\BPBI S.%
, {Bonifacio}, P.%
\ et al.\end{APACrefauthors}%
\unskip\
\newblock
\APACrefYearMonthDay{2018}{{\APACmonth{12}}}{},
\newblock
\unskip
\newblock
\APACjournalVolNumPages{\mnras}{481}{3}{3838-3852}.
\newblock
\begin{APACrefDOI} \doi{10.1093/mnras/sty2276} \end{APACrefDOI}
\PrintBackRefs{\CurrentBib}

\bibitem [\protect \citeauthoryear {%
{Starkenburg}%
\ \protect \BOthers {.}}{%
{Starkenburg}%
\ \protect \BOthers {.}}{%
{\protect \APACyear {2017}}%
}]{%
2017MNRAS.465.2212S}
\APACinsertmetastar {%
2017MNRAS.465.2212S}%
\begin{APACrefauthors}%
{Starkenburg}, E.%
, {Oman}, K\BPBI A.%
, {Navarro}, J\BPBI F.%
\ et al.\end{APACrefauthors}%
\unskip\
\newblock
\APACrefYearMonthDay{2017}{{\APACmonth{02}}}{},
\newblock
\unskip
\newblock
\APACjournalVolNumPages{\mnras}{465}{2}{2212-2224}.
\newblock
\begin{APACrefDOI} \doi{10.1093/mnras/stw2873} \end{APACrefDOI}
\PrintBackRefs{\CurrentBib}

\bibitem [\protect \citeauthoryear {%
{Taani}%
, {Abushattal}%
\BCBL {}\ \BBA {} {Mardini}%
}{%
{Taani}%
, {Abushattal}%
\BCBL {}\ \BBA {} {Mardini}%
}{%
{\protect \APACyear {2019}}%
}]{%
2019AN....340..847T}
\APACinsertmetastar {%
2019AN....340..847T}%
\begin{APACrefauthors}%
{Taani}, A.%
, {Abushattal}, A.%
\BCBL {}\ \BBA {} {Mardini}, M\BPBI K.%
\end{APACrefauthors}%
\unskip\
\newblock
\APACrefYearMonthDay{2019}{{\APACmonth{11}}}{},
\newblock
\unskip
\newblock
\APACjournalVolNumPages{Astronomische Nachrichten}{340}{847}{847-851}.
\newblock
\begin{APACrefDOI} \doi{10.1002/asna.201913713} \end{APACrefDOI}
\PrintBackRefs{\CurrentBib}

\bibitem [\protect \citeauthoryear {%
{Taani}%
, {Karino}%
\BCBL {}\ \BBA {} et al.%
}{%
{Taani}%
, {Karino}%
\BCBL {}\ \BBA {} et al.%
}{%
{\protect \APACyear {2019}}%
}]{%
2019JPhCS1258a2029T}
\APACinsertmetastar {%
2019JPhCS1258a2029T}%
\begin{APACrefauthors}%
{Taani}, A.%
, {Karino}, S.%
\BCBL {}\ \BBA {} et al., S.%
\end{APACrefauthors}%
\unskip\
\newblock
\APACrefYearMonthDay{2019}{{\APACmonth{10}}}{},
\newblock
{\BBOQ}\APACrefatitle {{On the wind accretion model of GX 301-2}} {{On the wind
  accretion model of GX 301-2}}.{\BBCQ}
\newblock
\BIn{} \APACrefbtitle {Journal of Physics Conference Series} {Journal of
  Physics Conference Series}\ \BVOL\ 1258, \BPG~012029.
\newblock
\begin{APACrefDOI} \doi{10.1088/1742-6596/1258/1/012029} \end{APACrefDOI}
\PrintBackRefs{\CurrentBib}

\bibitem [\protect \citeauthoryear {%
{The Astropy Collaboration}%
}{%
{The Astropy Collaboration}%
}{%
{\protect \APACyear {2013}}%
}]{%
Astropy13}
\APACinsertmetastar {%
Astropy13}%
\begin{APACrefauthors}%
{The Astropy Collaboration}.%
\end{APACrefauthors}%
\unskip\
\newblock
\APACrefYearMonthDay{2013}{}{},
\newblock
\unskip
\newblock
\APACjournalVolNumPages{A\&A}{558}{}{A33}.
\newblock
\begin{APACrefURL} \url{https://doi.org/10.1051/0004-6361/201322068}
  \end{APACrefURL}
\newblock
\begin{APACrefDOI} \doi{10.1051/0004-6361/201322068} \end{APACrefDOI}
\PrintBackRefs{\CurrentBib}

\bibitem [\protect \citeauthoryear {%
{Tissera}%
, {Beers}%
, {Carollo}%
\BCBL {}\ \BBA {} {Scannapieco}%
}{%
{Tissera}%
\ \protect \BOthers {.}}{%
{\protect \APACyear {2014}}%
}]{%
2014MNRAS.439.3128T}
\APACinsertmetastar {%
2014MNRAS.439.3128T}%
\begin{APACrefauthors}%
{Tissera}, P\BPBI B.%
, {Beers}, T\BPBI C.%
, {Carollo}, D.%
\BCBL {}\ \BBA {} {Scannapieco}, C.%
\end{APACrefauthors}%
\unskip\
\newblock
\APACrefYearMonthDay{2014}{{\APACmonth{04}}}{},
\newblock
\unskip
\newblock
\APACjournalVolNumPages{\mnras}{439}{3}{3128-3138}.
\newblock
\begin{APACrefDOI} \doi{10.1093/mnras/stu181} \end{APACrefDOI}
\PrintBackRefs{\CurrentBib}

\bibitem [\protect \citeauthoryear {%
{Vasiliev}%
}{%
{Vasiliev}%
}{%
{\protect \APACyear {2019}}%
}]{%
2019MNRAS.482.1525V}
\APACinsertmetastar {%
2019MNRAS.482.1525V}%
\begin{APACrefauthors}%
{Vasiliev}, E.%
\end{APACrefauthors}%
\unskip\
\newblock
\APACrefYearMonthDay{2019}{{\APACmonth{01}}}{},
\newblock
\unskip
\newblock
\APACjournalVolNumPages{\mnras}{482}{2}{1525-1544}.
\newblock
\begin{APACrefDOI} \doi{10.1093/mnras/sty2672} \end{APACrefDOI}
\PrintBackRefs{\CurrentBib}

\bibitem [\protect \citeauthoryear {%
{Yoneyama}%
}{%
{Yoneyama}%
}{%
{\protect \APACyear {1972}}%
}]{%
1972PASJ...24...87Y}
\APACinsertmetastar {%
1972PASJ...24...87Y}%
\begin{APACrefauthors}%
{Yoneyama}, T.%
\end{APACrefauthors}%
\unskip\
\newblock
\APACrefYearMonthDay{1972}{{\APACmonth{01}}}{},
\newblock
\unskip
\newblock
\APACjournalVolNumPages{\pasj}{24}{}{87}.
\PrintBackRefs{\CurrentBib}

\bibitem [\protect \citeauthoryear {%
{Yoon}%
, {Beers}%
, {Tian}%
\BCBL {}\ \BBA {} {Whitten}%
}{%
{Yoon}%
\ \protect \BOthers {.}}{%
{\protect \APACyear {2019}}%
}]{%
2019ApJ...878...97Y}
\APACinsertmetastar {%
2019ApJ...878...97Y}%
\begin{APACrefauthors}%
{Yoon}, J.%
, {Beers}, T\BPBI C.%
, {Tian}, D.%
\BCBL {}\ \BBA {} {Whitten}, D\BPBI D.%
\end{APACrefauthors}%
\unskip\
\newblock
\APACrefYearMonthDay{2019}{{\APACmonth{06}}}{},
\newblock
\unskip
\newblock
\APACjournalVolNumPages{\apj}{878}{2}{97}.
\newblock
\begin{APACrefDOI} \doi{10.3847/1538-4357/ab1ead} \end{APACrefDOI}
\PrintBackRefs{\CurrentBib}

\bibitem [\protect \citeauthoryear {%
{Yoon}%
\ \protect \BOthers {.}}{%
{Yoon}%
\ \protect \BOthers {.}}{%
{\protect \APACyear {2020}}%
}]{%
2020ApJ...894....7Y}
\APACinsertmetastar {%
2020ApJ...894....7Y}%
\begin{APACrefauthors}%
{Yoon}, J.%
, {Whitten}, D\BPBI D.%
, {Beers}, T\BPBI C.%
, {Lee}, Y\BPBI S.%
, {Masseron}, T.%
\BCBL {}\ \BBA {} {Placco}, V\BPBI M.%
\end{APACrefauthors}%
\unskip\
\newblock
\APACrefYearMonthDay{2020}{{\APACmonth{05}}}{},
\newblock
\unskip
\newblock
\APACjournalVolNumPages{\apj}{894}{1}{7}.
\newblock
\begin{APACrefDOI} \doi{10.3847/1538-4357/ab7daf} \end{APACrefDOI}
\PrintBackRefs{\CurrentBib}

\bibitem [\protect \citeauthoryear {%
{Yoshii}%
\ \BBA {} {Sabano}%
}{%
{Yoshii}%
\ \BBA {} {Sabano}%
}{%
{\protect \APACyear {1979}}%
}]{%
1979PASJ...31..505Y}
\APACinsertmetastar {%
1979PASJ...31..505Y}%
\begin{APACrefauthors}%
{Yoshii}, Y.%
\BCBT {}\ \BBA {} {Sabano}, Y.%
\end{APACrefauthors}%
\unskip\
\newblock
\APACrefYearMonthDay{1979}{{\APACmonth{01}}}{},
\newblock
\unskip
\newblock
\APACjournalVolNumPages{\pasj}{31}{}{505-522}.
\PrintBackRefs{\CurrentBib}

\bibitem [\protect \citeauthoryear {%
Zhao%
, Chen%
\BCBL {}\ \BBA {} et al.%
}{%
Zhao%
\ \protect \BOthers {.}}{%
{\protect \APACyear {2006}}%
}]{%
Zhao_2006}
\APACinsertmetastar {%
Zhao_2006}%
\begin{APACrefauthors}%
Zhao, G.%
, Chen, Y\BHBI Q.%
\BCBL {}\ \BBA {} et al., J\BHBI R\BPBI S.%
\end{APACrefauthors}%
\unskip\
\newblock
\APACrefYearMonthDay{2006}{jun}{},
\newblock
\unskip
\newblock
\APACjournalVolNumPages{Chinese Journal of Astronomy and
  Astrophysics}{6}{3}{265--280}.
\newblock
\begin{APACrefURL} \url{https://doi.org/10.1088%2F1009-9271%2F6%2F3%2F01}
  \end{APACrefURL}
\newblock
\begin{APACrefDOI} \doi{10.1088/1009-9271/6/3/01} \end{APACrefDOI}
\PrintBackRefs{\CurrentBib}

\bibitem [\protect \citeauthoryear {%
Zhao%
, Zhao%
, Chu%
, Jing%
\BCBL {}\ \BBA {} Deng%
}{%
Zhao%
\ \protect \BOthers {.}}{%
{\protect \APACyear {2012}}%
}]{%
Zhao_2012}
\APACinsertmetastar {%
Zhao_2012}%
\begin{APACrefauthors}%
Zhao, G.%
, Zhao, Y\BHBI H.%
, Chu, Y\BHBI Q.%
, Jing, Y\BHBI P.%
\BCBL {}\ \BBA {} Deng, L\BHBI C.%
\end{APACrefauthors}%
\unskip\
\newblock
\APACrefYearMonthDay{2012}{jun}{},
\newblock
\unskip
\newblock
\APACjournalVolNumPages{Research in Astronomy and
  Astrophysics}{12}{7}{723--734}.
\newblock
\begin{APACrefURL} \url{https://doi.org/10.1088%2F1674-4527%2F12%2F7%2F002}
  \end{APACrefURL}
\newblock
\begin{APACrefDOI} \doi{10.1088/1674-4527/12/7/002} \end{APACrefDOI}
\PrintBackRefs{\CurrentBib}

\bibitem [\protect \citeauthoryear {%
{Zolotov}%
\ \protect \BOthers {.}}{%
{Zolotov}%
\ \protect \BOthers {.}}{%
{\protect \APACyear {2009}}%
}]{%
2009ApJ...702.1058Z}
\APACinsertmetastar {%
2009ApJ...702.1058Z}%
\begin{APACrefauthors}%
{Zolotov}, A.%
, {Willman}, B.%
, {Brooks}, A\BPBI M.%
\ et al.\end{APACrefauthors}%
\unskip\
\newblock
\APACrefYearMonthDay{2009}{{\APACmonth{09}}}{},
\newblock
\unskip
\newblock
\APACjournalVolNumPages{\apj}{702}{2}{1058-1067}.
\newblock
\begin{APACrefDOI} \doi{10.1088/0004-637X/702/2/1058} \end{APACrefDOI}
\PrintBackRefs{\CurrentBib}

\end{thebibliography}

\end{document}